\documentclass[onecolumn,floatfix,superscriptaddress,a4paper,
               showpacs,showkeys,nofootinbib,preprint]{revtex4}
\usepackage{epsfig}
\usepackage{latexsym}
\usepackage{xspace}
\usepackage{hyperref}
\usepackage[latin2]{inputenc}
\usepackage{indentfirst}
\usepackage{enumerate}
\usepackage{color}

\usepackage{amsmath}
\usepackage{amssymb}
\usepackage[english]{babel}
\usepackage{url}

\newcommand{\eq}[1]{\begin{align} #1 \end{align}}

\begin{document}

\title{Non-Gaussian particle number fluctuations \\
in vicinity of the critical point
for van der Waals equation of state}
\author{V. Vovchenko}
\affiliation{
Taras Shevchenko National University of Kiev, 03022 Kiev, Ukraine}
\affiliation{
Frankfurt Institute for Advanced Studies, Johann Wolfgang Goethe University,
D-60438 Frankfurt, Germany}
\affiliation{
GSI Helmholtzzentrum f\"ur Schwerionenforschung GmbH, D-64291 Darmstadt, Germany}
\author{R.~V. Poberezhnyuk}
\affiliation{
Bogolyubov Institute for Theoretical Physics, 03680 Kiev, Ukraine}
\affiliation{
Frankfurt Institute for Advanced Studies, Johann Wolfgang Goethe University,
D-60438 Frankfurt, Germany}
\author{D.~V. Anchishkin}
\affiliation{
Bogolyubov Institute for Theoretical Physics, 03680 Kiev, Ukraine}
\affiliation{
Taras Shevchenko National University of Kiev, 03022 Kiev, Ukraine}
\affiliation{
Frankfurt Institute for Advanced Studies, Johann Wolfgang Goethe University,
D-60438 Frankfurt, Germany}
\author{M.~I. Gorenstein}
\affiliation{
Bogolyubov Institute for Theoretical Physics, 03680 Kiev, Ukraine}
\affiliation{
Frankfurt Institute for Advanced Studies, Johann Wolfgang Goethe University,
D-60438 Frankfurt, Germany}

\begin{abstract}

The non-Gaussian measures of the
particle number fluctuations -- skewness $S\sigma$ and kurtosis $\kappa \sigma^2$ --
are calculated in a vicinity of the critical point. This point corresponds to the end
point  of the first-order liquid-gas phase transition.
The gaseous phase is characterized by the positive values of
skewness while the liquid phase has negative skew.
The kurtosis appears to be
significantly negative at the critical density and supercritical temperatures.
The skewness and kurtosis  diverge at the critical point.
The classical van der Waals  equation of state in the
grand canonical ensemble formulation is used in our studies.
Neglecting effects of the
quantum statistics we succeed to obtain the
analytical expressions for the rich structures
of the skewness
and kurtosis in a wide region around the
critical point.
These results have universal form, i.e., they do not depend on particular values of the van der Waals parameters $a$ and $b$.
The strongly intensive measures
of particle number and energy fluctuations are also considered
and show singular behavior in the vicinity of the critical point.
\end{abstract}



\maketitle

\section{Introduction}
The first-order phase transition is among the most general phenomena in physics.
This phase transition exists in atomic and/or molecular
systems, in the system of interacting nucleons (nuclear matter), and,
most probably, in the QCD, between hadrons and quark-gluon plasma at large baryonic densities.
The liquid-gas phase transition line in the plane of temperature $T$ and chemical
potential $\mu$ has the end point, which is called the critical point (CP).
The CP demonstrates
some universal features typical for the second-order phase transitions, particularly,
anomalously large fluctuations.
The study of event-by-event  fluctuations in high-energy nucleus-nucleus
collisions opens new possibilities to investigate properties of strongly
interacting matter (see, e.g., Refs.~\cite{Koch:2008ia} and \cite{G-2015} and references therein) and the experimental search for the QCD CP is now in progress (see, e.g.,
Ref.~\cite{GGS-2014} and references therein).
The fluctuation signals of the QCD CP were discussed in Ref.~\cite{fluc3},
and higher moments of fluctuations of conserved charges were suggested as probes
to study the phase structure of QCD~\cite{Ejiri,Asakawa}.
Particularly,
the higher-order non-Gaussian measures such as the skewness $S \sigma$ and
kurtosis $\kappa \sigma^2$
of conserved charges fluctuations
have attracted much attention
(see, e.g., Ref.~\cite{Stephanov1} and \cite{Stephanov2}).
Experimentally, the STAR collaboration has measured the higher moments of net-proton and net-charge multiplicity in Au+Au collisions~\cite{STAR1,STAR2,STAR3}. 
See also recent review~\cite{Gaz-Seyb} 
on the search for critical behavior of strongly interacting 
matter at the CERN Super Proton Synchrotron.
Calculations of higher moments of conserved charges has been performed in various effective QCD models~\cite{Schaefer,Morita2013,Morita2012,Morita2014,Chatterjee2015}.
The effects of non-equilibrium evolution of these observables
in heavy-ion collisions
have also been considered recently~\cite{Mukherjee2015}.
Another fluctuation measures of interest are the so-called strongly
intensive quantities~\cite{SI} which are normally not sensitive to the fluctuations of the
system volume. This is especially relevant
for heavy-ion collision experiments where size of the
colliding system varies strongly on event-by-event level.

The van der Waals (VDW) equation is a simple analytical model
of the pressure function $p$ for equilibrium systems of particles with both attractive and
repulsive interactions. In the canonical ensemble (CE) it reads as
(see, e.g., Refs.~\cite{greiner,LL}),
\begin{equation}
p(T,n) ~=~ \frac{NT}{V-bN} ~-~ a \frac{N^2}{V^2}~ \equiv~\frac{n\,T}{1-bn}~-~a\,n^2~,
\label{eq:vdw}
\end{equation}
where $n\equiv N/V$ is the
particle number density while
the VDW parameters $a>0$ and $b>0$ describe the
attractive and repulsive interactions, respectively.
The first term on the right-hand-side of Eq.~\eqref{eq:vdw} corresponds
to the excluded volume (EV) correction, which manifests
itself in a substitution of a total volume $V$
by the available volume, $V_{\rm av} = V - b\,N$. The second
term comes from the mean field which describes attractive interactions between particles.
With regards to its asymptotic behavior in the vicinity of
the critical point the VDW model belongs to the mean-field theory universality class.
The number of particles $N$ is fixed in the CE.
In order to apply the VDW equation of state
to systems with variable number of particles
and calculate their fluctuations the grand canonical ensemble (GCE)
formulation is needed.
This procedure was firstly performed for the
EV model, i.e., for $a=0$ in Eq.~(\ref{eq:vdw}), in Refs.~\cite{vdw-1,vdw-2}.
In our recent paper \cite{VDW-GCE}, the full VDW equation \eqref{eq:vdw},
with both attractive and repulsive terms, was transformed
from the CE to the GCE for
systems with Boltzmann statistics while the
formulation which properly includes effects of quantum
statistics was obtained in Ref.~\cite{VDW-NM}.
Note that the EV and VDW models can also be conveniently treated within the GCE
in a framework of the thermodynamic mean-field approach
(see Refs.~\cite{mf-1992,mf-1995,mf-2014}).

In the present paper we use our recent results of the GCE formulation~\cite{VDW-GCE}
as a starting point for calculating the scaled variance, skewness, and kurtosis
of particle number fluctuations, and also the strongly intensive
measures of particle number and energy fluctuations.
The Boltzmann approximation is adopted.
This gives a possibility to obtain the analytical expressions for the
universal structure of fluctuations in a vicinity of the CP.

The paper is organized as follows.
In Sec.~\ref{sec-vdw} the VDW equation of state is formulated within the GCE.
In Sec.~\ref{sec-fluct} the particle number fluctuations -- scaled variance,
skewness, and kurtosis~-- are calculated and their behavior
in a vicinity of the CP is analyzed.
In Sec.~\ref{sec-si} the strongly intensive measures
of fluctuations for the energy and number of particles  are considered.
A summary in Sec.~\ref{sec-sum} closes the article.

\section{VDW equation
in the GCE}
\label{sec-vdw}
The canonical ensemble VDW pressure function in (1) corresponds to the Boltzmann approximation. This expression is used in
the present paper, i.e., effects of the quantum statistics (Bose or Fermi)
will be neglected.
The VDW pressure is a unique function of variables $T$ and $n$
for all $T\ge 0$ and $0\le n\le 1/b$, and
this equation of state contains the first-order liquid-gas phase transition
and has the CP.
The CP in $(T,n)$-plane, i.e. the point $(T_c,n_c)$, corresponds to the temperature
and particle number density, where the following derivatives are equal to zero,
\begin{equation}
\left(\frac {\partial p}{\partial n}\right)_T~=~0~,
~~~~~~\left(\frac {\partial ^2 p}{\partial n^2}\right)_T~=~0~.
\label{p-der}
\end{equation}
and the thermodynamical quantities at the CP are equal to:
\eq{\label{crit}
 T_c = \frac{8a}{27b}~,~~~~~ n_c =
\frac{1}{3b}~,~~~~~ p_c = \frac{a}{27b^2}~.
}
At $T>T_c$ the following inequality is always valid,
\eq{\label{p-der-1}
\left(\frac {\partial p}{\partial n}\right)_T~>~0~,
}
while at $T<T_c$ there appears an unstable interval $[n_1,n_2]$ with
\eq{\label{p-der-2}
\left(\frac {\partial p}{\partial n}\right)_T~<~0~.
}
This means that the VDW isotherm $p(T,n)$ has a local maximum at $n=n_1$ and a local minimum at $n=n_2>n_1$ for $T<T_c$.
The unstable part (\ref{p-der-2}) of the VDW isotherm at the interval
$[n_1,n_2]$, together
with two additional metastable parts -- $[n_g,n_1]$ and $[n_2,n_l]$ --
are transformed to a mixture of two phases: a gas with density $n_g<n_1$
and a liquid with density $n_l>n_2$. This is done in accordance with the Maxwell rule
of the equal
areas (see, e.g., Refs.~\cite{greiner,LL}) which leads to a constant pressure $p(T,n_g)=p(T,n_l)$
inside the density interval $[n_g,n_l]$.

In the GCE the pressure should be defined in terms of its natural
variables: temperature $T$ and chemical potential $\mu$.
The $p(T,\mu)$ function contains a complete information about equilibrium physical systems.
Other thermodynamical quantities, such as
particle number density $n(T,\mu)$, entropy density $s(T,\mu)$,
and energy density $\varepsilon(T,\mu)$ can be presented
in terms of $p$ and its $T$- and $\mu$-derivatives:
\begin{equation}
n(T,\mu)=\left(\frac{\partial p}{\partial \mu}\right)_T~,~~~~
s(T,\mu)=\left(\frac{\partial p}{\partial T}\right)_{\mu}~,~~~~\varepsilon (T,\mu)=
T \left(\frac{\partial p}{\partial T}\right)_{\mu}+
\mu \left(\frac{\partial p}{\partial \mu}\right)_{T}-p~.
\label{pTmu-s}
\end{equation}

The VDW equation of state in the GCE is obtained in the
form of transcendental equation for particle number density
$n \equiv n(T,\mu)$ as a function of $T$ and $\mu$ (see Ref. \cite{VDW-GCE} for details):
\begin{equation}
n(T,\mu) ~ =~ \frac{n^{\rm id}(T, \mu^*)}{1~+~b \,n^{\rm id}(T, \mu^*)}~,~~~~~
\mu^* ~=~ \mu~ -~ T \frac{bn}{1\,-\,bn} ~+~ 2 a n~,
\label{eq:nvdwtr}
\end{equation}
where $n^{\rm id}$ is a particle number density in the ideal  Boltzmann gas
\begin{equation}
n^{\rm id}(T, \mu) ~ =~
\exp\left(\frac{\mu}{T}\right)\,\frac{d\,m^2\, T}{2 \pi^2} \, K_2\left(\frac{m}{T}\right)
\label{n-id}
\end{equation}
with $d$ being the degeneracy factor and $m$ being the particle mass.
The $K_2(x)$ is the modified Bessel function of the second kind.
The GCE VDW pressure $p(T,\mu)$ is then obtained by inserting $n(T,\mu)$ from
(\ref{eq:nvdwtr}) into Eq.~(\ref{eq:vdw}).
Note, the relativistic form of a dispersion relation is
considered, $\omega(k)=\sqrt{m^2+k^2}$, where $\omega$ and $k$ are
the free single-particle energy and momentum, respectively\footnote{We use
the system of units, where the Plank constant $\hbar$, the speed of
light $c$, and the Boltzmann constant $k$ are equal to unity,
$\hbar=c=k=1$.}.
This makes the present formulation being suitable for high-energy physics
applications.

In the GCE there is a unique solution of Eq.~\eqref{eq:nvdwtr} at $T>T_c$, while
at $T<T_c$ it
may have either one solution or three different solutions for particle number density
$n(T,\mu)$.
In that case
the solution which corresponds to a largest pressure survives
in accordance to the Gibbs criterion.
The liquid-gas mixed phase in the $T$-$\mu$ plane belongs to the
line $\mu=\mu_c(T)$, where two solutions  with different particle number densities,
$n_g(T,\mu)$ and $n_l(T,\mu)$, correspond to the equal pressures,
$p_g(T,\mu)=p_l(T,\mu)$. The Maxwell rule
of the equal areas and the Gibbs criteria
of equal pressures for the gas and liquid at
the same $T$ and $\mu$ values appear to be
the equivalent
descriptions of the first-order liquid-gas phase
transitions (see Ref.~\cite{VDW-NM} for details).

\section{Particle number fluctuations}
\label{sec-fluct}

\subsection{Fluctuations in the GCE}
Let the particle number $N$ be a random variable
with the normalized  probability distribution ${\cal P}(N)$.
The $k$-th moment $\langle N^k \rangle$ is then defined as
\eq{
\langle N^k \rangle = \sum_N N^k {\cal P}(N)~.
}
Let us introduce the variance,
$\sigma^2 = \langle (\Delta N)^2 \rangle$, where $\Delta N\equiv N-\langle N\rangle$.
The scaled variance,
\eq{\label{omega}
\omega[N]~\equiv~\frac{\sigma^2}{\langle N\rangle} ~,
}
characterizes the width of the ${\cal P}(N)$ distribution. Note that
$\omega[N]=1$ for the Poisson distribution
${\cal P}(N)=\exp(-\langle N\rangle)\,\langle N\rangle^N/N!$.

The skewness $S\sigma$
is defined as
\eq{\label{skew}
S\sigma ~=~ \frac{\langle (\Delta N)^3 \rangle}{\sigma^2}~.
}
The skewness measures the degree of asymmetry of the distribution
${\cal P}(N)$ around its mean
value $\langle N\rangle$.
Positive skewness indicates a distribution with an asymmetric tail
extending more to the {\it right}, i.e., toward  $N$-values with $N>\langle N\rangle$.
Negative skewness indicates a distribution with an asymmetric tail
extending more to the {\it left}, i.e., toward  $N$-values with $N<\langle N\rangle$.
If the ${\cal P}(N)$ distribution is symmetric around its mean value,
i.e., the {\it right} and {\it left} tails are equal, it has zero skewness.
This is the case for the normal Gaussian distribution, whereas the Poisson distribution
shows a positive value of the skewness, $S\sigma=1$.

The (excess) kurtosis $\kappa\sigma^2$ is the measure of ``peakedness'' of the probability
distribution ${\cal P}(N)$,
\eq{\label{kurt}
\kappa\sigma^2~ =~ \frac{\langle (\Delta N)^4 \rangle ~-~ 3\, \langle (\Delta N)^2 \rangle^2 }{\sigma^2}~.
}
The kurtosis (\ref{kurt}) measures the degree to which a distribution is
more or less peaked than a normal Gaussian distribution. Positive kurtosis indicates a
relatively peaked distribution. Negative kurtosis indicates a relatively flat distribution.
For the Poisson distribution
one has positive value of the kurtosis, $\kappa \sigma^2 = 1$.

The normal Gaussian distribution corresponds to the zero value of both
the skewness (\ref{skew}) and the (excess) kurtosis (\ref{kurt}). Therefore, (strong)
deviations of $S\sigma$ and/or $\kappa\sigma^2$ from zero  are the signatures
of the (highly) non-Gaussian shape of the particle number distribution ${\cal P}(N)$.

In the GCE the system is defined by the pressure $p$ given in terms of its natural variables $T$ and $\mu$.
The particle number fluctuations can be characterized by the following dimensionless
cumulants (susceptibilities),
\eq{\label{kn}
k_n~=~\frac{\partial^n (p/T^4)}{\partial(\mu/T)^n}~,
}
which are connected to the moments of the particle number
distribution as
\eq{\label{kN}
k_1=\frac{\langle N\rangle}{VT^3} ~,~~~~ k_2=\frac{\langle (\Delta N)^2\rangle}
{VT^3}~,~~~~ k_3=\frac{\langle (\Delta N)^3\rangle}{VT^3}~,~~~~
k_4=\frac{\langle (\Delta N)^4\rangle~-~3\langle (\Delta N)^2\rangle^2}{VT^3}~,
}
where
$\langle \ldots\rangle$
denotes the GCE averaging.
The scaled variance (\ref{omega}), skewness (\ref{skew}), and kurtosis (\ref{kurt})
are the intensive fluctuation measures
that remain finite
in the thermodynamic limit $V \to \infty$.
They can be expressed in terms of the susceptibilities as the following
\eq{\label{osk}
\omega[N]~=~\frac{k_2}{k_1}~,~~~~~~~
S\sigma~=~\frac{k_3}{k_2}~,~~~~~~~ \kappa \sigma^2=\frac{k_4}{k_2}~.
}
%

\subsection{Scaled variance}
\begin{figure}[!h]
\begin{minipage}{.7\textwidth}
\center
\includegraphics[width=\textwidth]{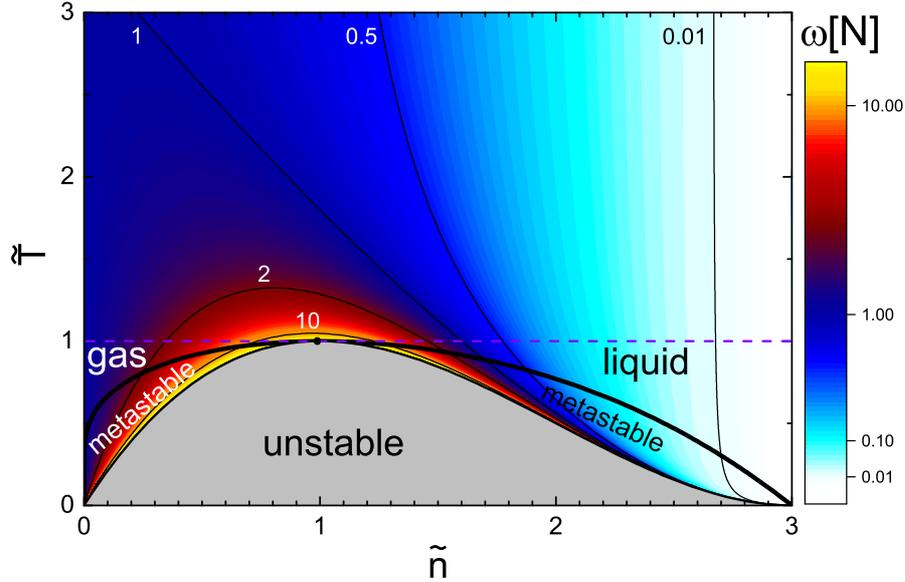}
\end{minipage}
\caption{(Color online)
The scaled variance $\omega[N]$ (\ref{omega-N})
on the $(\widetilde{n},\widetilde{T})$ phase diagram for
both stable and metastable pure phases.
Several lines of constant values of $\omega[N]$ are shown.
The grey area depicts region where pure phase is mechanically unstable.
} \label{fluct-ms}
\end{figure}

The scaled variance defined in Eq.~\eqref{osk} can be calculated through
the $\mu$-derivative of the particle density. For
pure phases in
the classical VDW gas
it results in (see also Ref.~\cite{VDW-GCE})
\begin{equation}
\omega[N]~ = ~\frac{T}{n} \, \left(\frac{\partial n}{\partial \mu}\right)_T
~=~
\left[\frac{1}{(1-bn)^2}-\frac{2a  n}{T}\right]^{-1}~.
\label{omega-N}
\end{equation}
In terms of the reduced quantities, $\widetilde{T}\equiv T/T_c$ and
$\widetilde{n} \equiv n/n_c$,
Eq.(\ref{omega-N}) reads as
\begin{equation}
\omega[N]=\frac{1}{9}\left[\frac{1}{(3-\widetilde{n})^2}-\frac{\widetilde{n}}
{4 \widetilde{T}}\right]^{-1}~,
\label{omega-N1}
\end{equation}
and possesses
the universal form independent of the specific values of the VDW parameters $a$ and $b$.
From Eq.~(\ref{omega-N1}) it follows that $\omega[N]\rightarrow 1$ at $\widetilde{n}\rightarrow 0$
(this corresponds to the ideal gas limit and the Poisson ${\cal P}(N)$ distribution),
and $\omega[N]\rightarrow 0$
at $\widetilde{n}\rightarrow 3$ (this corresponds to the liquid with the highest possible density).
According to its definition, the scaled variance $\omega[N]$ is a positive quantity.
This is indeed the case for all $T$ and $n$ values that correspond to stable
and even metastable states. The scaled variance diverges, $\omega[N]\rightarrow \infty$,
at the CP.
Introducing quantities $\rho=\widetilde{n}-1$ and
$\tau=\widetilde{T}-1$ one finds at $\tau\ll 1$ and $\rho\ll 1$:
\eq{
\omega[N]\cong\frac{4}{9}\left[\tau+\frac{3}{4}\rho^2+\tau\rho\right]^{-1}~.
}
The scaled variance (\ref{omega-N}) as a function of $\widetilde{T}$ and $\widetilde{n}$,
is plotted
in Fig.~\ref{fluct-ms}
for both stable and metastable pure phases.

\subsection{Skewness}
The skewness $S\sigma$ can be calculated as
\eq{\label{skew1}
S\sigma ~= ~\frac{k_3}{k_2}~=~ \omega[N] + \frac{T}{\omega[N]} \,
\left(\frac{\partial \omega[N]}{\partial \mu}\right)_{T}~=~
(\omega[N])^2\left[\frac{1-3 bn}{(1-bn)^3}\right]~.
}
In the reduced variables it reads
\eq{\label{eq:sN}
S\sigma=\frac{1}{3}\left[\frac{1}{(3-\widetilde{n})^2}-\frac{\widetilde{n}}{4 \widetilde{T}}\right]^{-2}~
\left[\frac{1-\widetilde{n}}{(3-\widetilde{n})^3}\right]~.
}
%

\begin{figure}[!h]
\begin{minipage}{.8\textwidth}
\centering
\includegraphics[width=\textwidth]{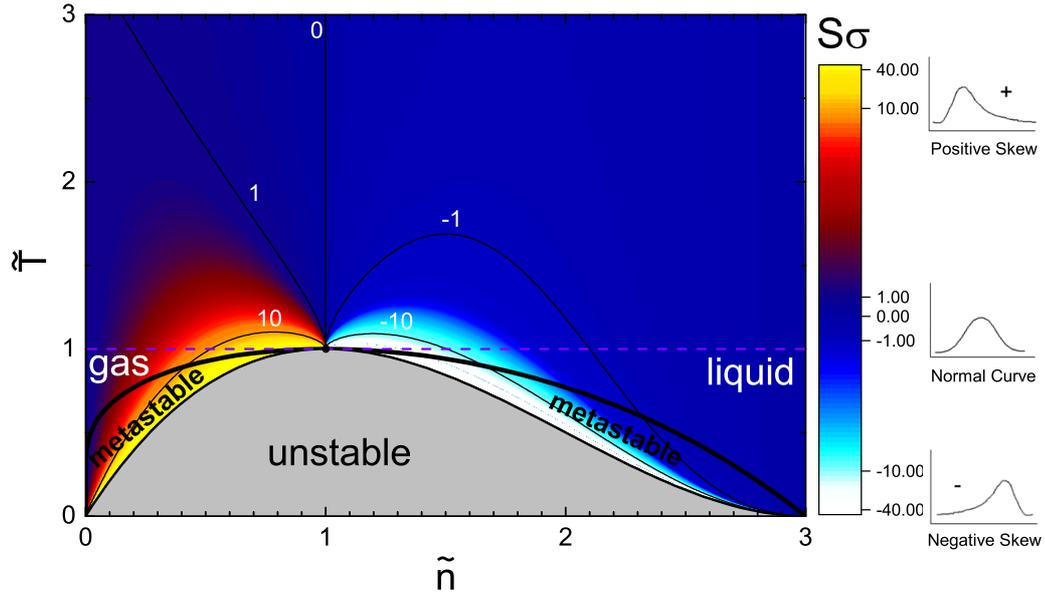}
\end{minipage}
\caption{(Color online)
The same as in Fig.~\ref{fluct-ms}
but for the skewness $S\sigma$~\eqref{eq:sN}.
Different values of skewness are illustrated by the typical
${\cal P}(N)$ distributions on the right panel.
} \label{fluct-s-ms}
\end{figure}

As a function of the reduced temperature and density the skewness
$S\sigma$ is plotted
in Fig.~\ref{fluct-s-ms}
for both stable and metastable pure phases.
It is clearly seen from Eqs.~(\ref{skew1}) and (\ref{eq:sN}) that the skewness is positive
at $\widetilde{n}<1$ (the gaseous phase),  negative
at $\widetilde{n}>1$ (the liquid phase), and $S\sigma=0$ at $\widetilde{n}=1$.
At $\widetilde{n}\rightarrow 0$ one finds that $S \sigma \to 1$.
This is a small asymmetry of the particle number distribution and
it corresponds to the Poisson distribution that takes place
for the ideal Boltzmann gas.

In a vicinity of the CP
one finds,
\eq{\label{skew2}
S\sigma\cong -~\frac{2}{3}\rho\,\left[\tau+\frac{3}{4}\rho^2+\tau\rho\right]^{-2}~.
}
The CP, $\widetilde{T}=\widetilde{n}=1$, i.e., $\rho=\tau=0$, is a point of
the essential singularity
of the skewness measure. For example, at $\tau=0$ one finds from Eq.~(\ref{skew2}) that
$S\sigma\rightarrow +\,\infty$ at $\rho\rightarrow +\,0$ and
$S\sigma\rightarrow -\,\infty$ at $\rho\rightarrow -\,0$. At the same time,
$S\sigma=0$ at any $\tau>0$ and $\rho=0$.

\subsection{Kurtosis}
The kurtosis $\kappa\sigma^2$ can be calculated as
\eq{\label{eq:kN}
\kappa \sigma^2 ~=~ \frac{k_4}{k_2} ~=~ (S\sigma)^2 + T \,
\left(\frac{\partial [S\sigma]}{\partial \mu}\right)_{T}~=~
3\, (S\sigma)^2 - 2\, \omega[N] \, S\sigma-
54\, (\omega[N])^3\frac{\widetilde{n}^2}{(3-\widetilde{n})^4}~.
}
%
%
\begin{figure}[!h]
\begin{minipage}{.8\textwidth}
\centering
\includegraphics[width=\textwidth]{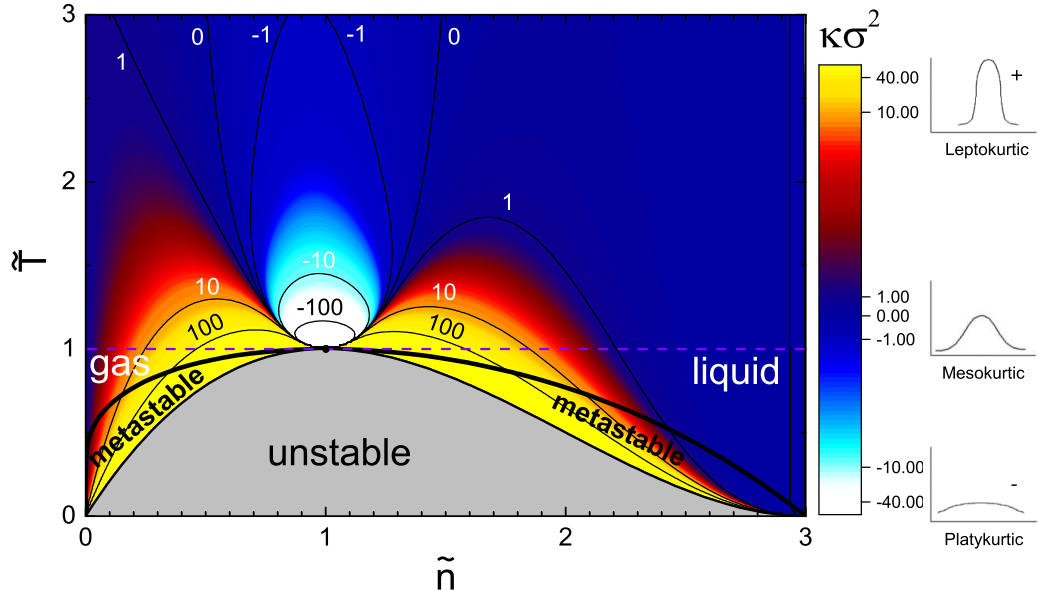}
\end{minipage}
\caption{(Color online)
The same as in Figs.~\ref{fluct-ms} and \ref{fluct-s-ms}
but for
the kurtosis $\kappa \sigma^2$~\eqref{eq:kN}.
Different values of kurtosis are illustrated by the typical
${\cal P}(N)$ distributions on the right panel.
} \label{fluct-k-ms}
\end{figure}
As a function of the reduced temperature and density the kurtosis
$\kappa \sigma^2$ is plotted
in Fig.~\ref{fluct-k-ms}
for both stable and metastable pure phases.
It is seen from these figures that at $\widetilde{T}<1$
the kurtosis is positive (leptokurtic)
for both $\widetilde{n}<1$ (the gaseous phase) and $\widetilde{n}>1$
(the liquid phases).

Approaching the CP the kurtosis diverges.
At $\tau=0$ and $\rho\ll 1$ one finds
\eq{\label{kurt-tau0}
\kappa \sigma^2~\propto~\rho^{-6}~.
}
Notably, the kurtosis has attains large negative values (platykurtic)
at critical density $\widetilde{n}=1$ and temperatures just above the critical, $\widetilde{T}>1$.
In this region, one finds
\eq{\label{kurt-rho0}
\kappa \sigma^2~\propto~-~\tau^{-3}~.
}
This indicates that particle number
distribution has a flat peak in that region, much
flatter than the corresponding Gaussian with the same width.
This region can be identified as the crossover region,
where rapid although smooth transition between gaseous and liquid phases takes place.
This is in line with arguments that the crossover region near the CP
is characterized by the negative sign of kurtosis~\cite{Stephanov2}.
In the vicinity of the CP
the kurtosis changes rapidly and attains both positive and negative values.

At $\widetilde{n}\rightarrow 0$ the VDW equation of state corresponds to the ideal Boltzmann gas
limit. In this case, ${\cal P}(N)$ approaches the Poisson  distribution, and, as it follows
from Eqs.~(\ref{eq:sN}) and (\ref{eq:kN}), the skewness and kurtosis both
approach the Poisson expectation values, i.e., $S \sigma \rightarrow 1$ and $\kappa \sigma^2 \rightarrow 1$.
It should be noted that the ideal Boltzmann gas with the Poisson
distribution reveals small deviations from the Gaussian ${\cal P}(N)$ distribution for which, by construction,
$S \sigma =0$ and $\kappa \sigma^2 =0$.

The nucleon number fluctuations
were recently considered in Ref.~\cite{VAGP} within the modified
VDW equation. To describe the nuclear matter ground state the  Fermi statistics
of nucleons was introduced \cite{VDW-NM}.
The Boltzmann approximation is assumed in the present study
and it corresponds to the original version of the VDW
equation.
This gives a possibility to obtain the  analytical expressions for the
scaled variance, skewness, and kurtosis.
Note also that our results for the particle number fluctuations
presented in Figs.~\ref{fluct-ms}-\ref{fluct-k-ms} are universal, i.e.,
they are independent on the specific numerical values of the VDW parameters $a$ and $b$.
Therefore, they can be applied to very different physical systems -- from the CP of
water with $T_c=647$~K$^0$ up to the CP in of the nuclear matter with $T_c\cong2\times 10^{12}$~K$^0$.
The universality of the results is lost if one would take into account the effects of quantum statistics.
On the other hand, in most cases the inclusion of Fermi statistics
changes the results on a quantitative level,
qualitatively they remain essentially the same as in case of the Boltzmann statistics.

\section{Strongly intensive quantities}
\label{sec-si}
The results of the previous section demonstrate
a strong increase of the particle number fluctuations in a vicinity
of the CP. The fluctuations may become also very large for metastable
states, i.e.,  super-cooled gas and/or super-heated liquid. These fluctuation signals
from phase transitions in the nuclear matter
can be observed in the event-by-event analysis of heavy ion collisions.
Note, however, that in these processes there is one more source of particle number fluctuations,
namely, event-by-event fluctuations of the system volume. These
volume fluctuations are mainly of the geometrical
origin, and they can hardly be avoided in nucleus-nucleus reactions.
Thus, one may observe large experimental fluctuations even for
simple statistical systems, e.g., the ideal gas.
The strongly intensive measures
of the fluctuations defined in terms of two extensive quantities $A$ and $B$
were suggested in Ref.~\cite{SI}.
For statistical systems in a case of the absence of phase transitions
these measures within the GCE formulation are independent  of the system volume and its fluctuations.
Note, however, that in systems with the CP (in general, for the 2nd order phase transitions)
the critical behavior does depend on the system volume and shows the characteristic finite-size scaling.
This implies that strongly intensive quantities are also volume-dependent near the CP. Thus, using the strongly intensive measures one excludes trivial volume fluctuations for normal statistical systems, and a presence of large fluctuations in terms of these measures can be considered as an indication of critical behavior.

In the present paper we consider the strongly intensive measures
of total energy $E$ and particle number $N$ fluctuations for the VDW equation of state.
They are defined as
\eq{\label{D}
\Delta[E,N]~ &= ~C_{\Delta}^{-1}\Big[\langle N \rangle \omega[E]-\langle E \rangle \omega[N] \Big]~,\\
\Sigma[E,N]~ &= ~C_{\Sigma}^{-1}\Big[\langle N \rangle \omega[E]+\langle E \rangle \omega[N]
-2 \Big(\langle EN\rangle -\langle E\rangle\,\langle N\rangle\Big)\Big]~,\label{S}
}
where $C_{\Delta}^{-1}$ and $C_{\Sigma}^{-1}$ are the normalization factors
that have been suggested in the following form
\cite{SI-Norm}
\eq{
C_{\Delta} ~=~ C_{\Sigma} ~=~ \langle N \rangle \, \omega[\varepsilon]~,
}
with $\omega[\varepsilon]$ being the scaled variance of
a single-particle energy distribution in the VDW system.
To proceed it is necessary to calculate $\omega[\varepsilon]$, $\omega[E]$, and
$\langle EN\rangle$.

In the VDW gas the average single-particle energy $\overline{\epsilon}$ is
independent of the parameter $b$, but it is modified due to a presence of the
attractive mean field:
\eq{
\overline{\epsilon}=\overline{\epsilon}_{\rm{id}}(T)-a\frac{N}{V}~=~
3~T+m\frac{K_1(m/T)}{K_2(m/T)}~-~a\,n~,
}
where $\overline{\epsilon}_{\rm id}$ is the average single-particle energy in
the relativistic ideal gas.
The variance of the single-particle energy is insensitive to the presence of the VDW mean field, and one obtains

\eq{\label{var-epsilon}
\omega[\epsilon]~=~
\frac{\overline{\epsilon^2}~-~\overline{\epsilon}^2}
{\overline{\epsilon}}~=~
\frac{T^2}{\overline{\epsilon}}~\frac{\partial\, \overline{\epsilon}_{\rm{id}}}{\partial T} ~.
}

The mean total energy is
\eq{\label{Mean-E}
\langle E \rangle~=~\left\langle \left(\overline{\epsilon}_{\rm{id}}-
a \frac{N}{V}\right)N\right\rangle\,
=\, \overline{\epsilon}_{\rm{id}}\,\langle N \rangle -
\frac{a}{V}\,\langle N^2 \rangle=\overline{\epsilon}_{\rm{id}}\,\langle N \rangle -
\frac{a}{V}\,\langle N \rangle^2- a\, \frac{\langle N^2\rangle -\langle N\rangle^2}{V}~.
}
The first and second terms in the right hand side of Eq.~(\ref{Mean-E})
are proportional to $\langle N\rangle$. On the other hand,
the third term remains finite outside the critical point
in the thermodynamic limit $V\rightarrow \infty$. Therefore, one obtains
\eq{\label{mean-E}
\langle E \rangle ~\cong ~(\overline{\epsilon}_{\rm{id}}-an)\,\langle N \rangle~.
}
For $\omega[E]$ one then finds
\eq{\label{omega-e}
\omega[E]\equiv \frac{\langle E^2\rangle -\langle E\rangle^2}{\langle E\rangle}
=\frac{1}{\langle E \rangle}\,T^2 \left(\frac{\partial \langle E \rangle}{\partial T}\right)_{\mu/T}
=\omega[\epsilon]+
\frac{(\overline{\epsilon}_{\rm{id}}-2 a n)^2}{\overline{\epsilon}_{\rm{id}}-a n}\,\omega[N]~.
}
Finally, the correlations between $E$ and $N$
can be calculated as the following
\eq{
\langle EN\rangle -\langle E\rangle\,\langle N\rangle
~=~ T^2 \left(\frac{\partial \langle N \rangle}{\partial T}\right)_{\mu/T}
= (\overline{\epsilon}_{\rm{id}}-2an) \, \langle N \rangle \, \omega[N]~.
}

Substituting the above formulae into Eqs.~(\ref{D}) and (\ref{S}) one finds
the following expressions for the strongly intensive quantities:
\eq{\label{D-1}
\Delta[E,N]~ &= ~1~-~\frac{a n(2\overline{\epsilon}_{\rm{id}}-3an)}
{\overline{\epsilon_{\rm id}^2}-\overline{\epsilon}_{\rm id}^2} ~\omega[N], \\
\Sigma[E,N]~ &= ~1+\frac{a^2 n^2}
{\overline{\epsilon_{\rm id}^2}-\overline{\epsilon}_{\rm id}^2} ~\omega[N]~.\label{S-1}
}
In the absence of
the attractive interactions (i.e., $a=0$),
one can readily see from Eqs.~(\ref{D-1}) and (\ref{S-1}) that
$\Delta[E,N] = \Sigma[E,N] = 1$, thus, in the excluded-volume
model the strongly intensive quantities are the same as in the ideal Boltzmann gas.

\begin{figure}[!h]
\begin{minipage}{.7\textwidth}
\centering
\includegraphics[width=\textwidth]{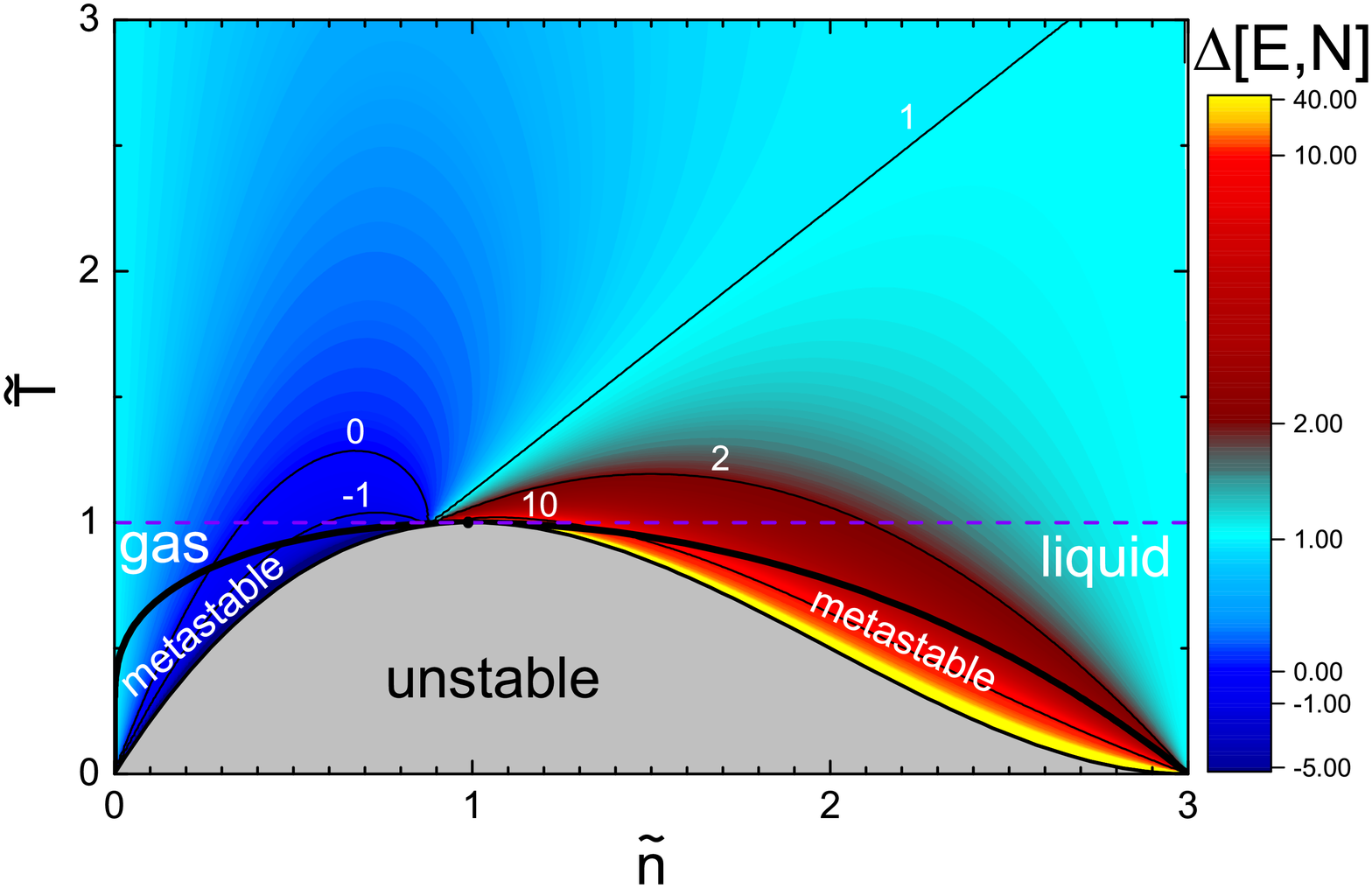}
\end{minipage}
\caption{(Color online)
The strongly intensive quantity $\Delta[E,N]$~\eqref{D-2}
on the $(\widetilde{n},\widetilde{T})$ phase diagram for
both stable and metastable pure phases.
Several lines of constant values of $\Delta[E,N]$
are shown.
} \label{fig:deltaEN}
\end{figure}

\begin{figure}[!h]
\begin{minipage}{.7\textwidth}
\centering
\includegraphics[width=\textwidth]{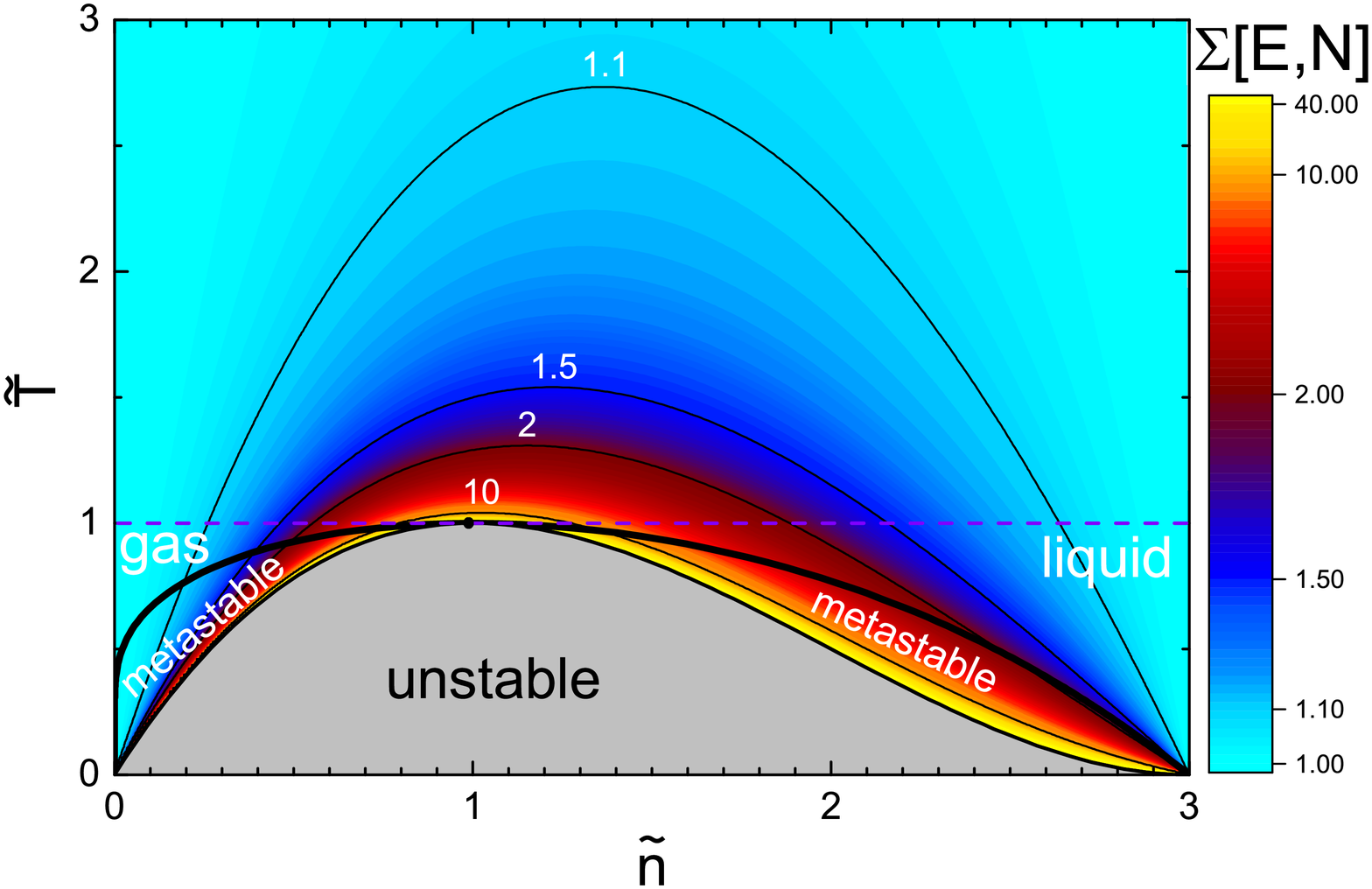}
\end{minipage}
\caption{(Color online)
The strongly intensive quantity $\Sigma[E,N]$~\eqref{S-2}
on the $(\widetilde{n},\widetilde{T})$ phase diagram for
both stable and metastable pure phases.
Several lines of constant values of $\Sigma[E,N]$
are shown.
} \label{fig:sigmaEN}
\end{figure}

The expressions (\ref{D-1}) and (\ref{S-1}) for $\Delta[E,N]$  and $\Sigma[E,N]$ become
more transparent if one considers the non-relativistic limit,
$\overline{\epsilon}_{\rm{id}} = 3T/2$ and
$\overline{\epsilon_{\rm id}^2}-\overline{\epsilon}_{\rm id}^2 = 3T^2/2$.
Note that the rest energy, $m$, has been subtracted, thus,
only the kinetic energy fluctuations contribute to
$\overline{\epsilon^2_{\rm id}}$.
One then obtains
\eq{
\Delta[E,N] &= 1 - \frac{2}{3} \frac{a n(3T - 3an)}{T^2} ~\omega[N] ~=~
~1~-~
\frac{9\widetilde{n}}{4\widetilde{T}}\, \left[1 ~-~
\frac{9\widetilde{n}}{8\widetilde{T}}  \right]\,\omega[N]~, \label{D-2} \\
\Sigma[E,N] &= 1 + \frac{2}{3} \frac{a^2 n^2}{T^2} ~\omega[N]~=~
1 + \frac{27}{32} \, \frac{\widetilde{n}^2}{\widetilde{T}^2} \, \omega[N]~.\label{S-2}
}
The quantities $\Delta[E,N]$ and $\Sigma[E,N]$
are depicted in Figs.~\ref{fig:deltaEN} and \ref{fig:sigmaEN}, respectively.
Both the $\Sigma[E,N]$ and  $\Delta[E,N]$ measures approach unity in both zero density,
$\widetilde{n} \to 0$, and packing, $\widetilde{n} \to 3$, limits and diverge at the CP.
Note that the $\Sigma[E,N]$ measure is always positive and even larger than unity,
while the $\Delta[E,N]$ measure
attains  both positive and negative values.

\section{Summary}
\label{sec-sum}

Particle number fluctuations up to the fourth order -- scaled variance $\omega[N]$, skewness
$S\sigma$, and kurtosis $\kappa\sigma^2$ -- have been calculated for the classical van der Waals equation of state. Analytical formulae
were obtained and used for an analysis of
the fluctuation behavior in a vicinity of the CP.
The obtained results have the universal form,
i.e., they are independent of the specific numerical values of the VDW parameters $a$ and $b$.

The skewness $S\sigma$ is positive at $n<n_c$ and negative at $n>n_c$
for all values of temperature. This means that gaseous and liquid phases
at $T<T_c$ are clearly characterized by positive and negative
values of $S\sigma$, respectively. Above the critical temperature $T_c$ the
skewness is equal to zero at the $n=n_c$ line, and this line can be associated
with the crossover transition from gaseous to liquid matter.

The kurtosis $\kappa \sigma^2$ is very sensitive to the proximity of
the critical point and also has a rich structure. At $T<T_c$ the
kurtosis is positive in both phases. In the vicinity of the CP
the kurtosis changes rapidly and can attain both positive and negative
values at $T>T_c$. Notably, in the crossover region just above $T_c$ it attains large negative values.

The strongly intensive measures $\Delta[E,N]$ and $\Sigma[E,N]$ for the
fluctuations of system energy and number of particles
have been also calculated.
In the excluded-volume model, i.e. at $a=0$, these measures are the same as in the ideal gas and are equal to unity.
In the full van der Waals equation, however, they diverge at the
critical point and differ significantly from unity in the region around it.
These measures can be used to study the fluctuations in the
nuclear matter created in heavy ion collisions.

\begin{acknowledgments}
We are thankful
to M. Ga\'zdzicki for fruitful comments and discussions.
This work was supported by the Humboldt
Foundation, by the Program of Fundamental Research
of the Department of Physics and Astronomy of National Academy of Sciences of Ukraine,
and by HIC for FAIR within the LOEWE program of the State of Hesse.
\end{acknowledgments}

\end{document}